# Drastic ground state changes induced by Ni substitution in $Na_xCoO_2$

N. Gayathri, M. Premila, A. Bharathi*, V. S. Sastry, C. S. Sundar and Y. Hariharan
Materials Science Division, Indira Gandhi Center for Atomic Research, Kalpakkam. 603 102. India.

Abstract

We report on the effect of Ni substitution at the Co site on the physical properties of $Na_xCoO_2$ system by investigating the series $Na_xCo_{1-y}Ni_yO_2$ (x=0.75, $0\leq y\leq 0.15$). An upturn in the resistivity is observed in all Ni substituted samples as the temperature is lowered, suggestive of the occurrence of a Metal-Insulator Transition (MIT). The temperature at which this transition occurs increases with Ni content. The temperature dependence of the resistivity in the metallic region in the Ni substituted samples shows a $T^2$ dependence, which is qualitatively different from that observed in the pristine sample. The evolution of the Fano asymmetry parameter, extracted by analyzing the lineshape of the IR active in-plane Co-O mode, both as a function of Ni concentration and temperature corroborates the occurrence of the MIT. It is argued that the progressive substitution of the $Co^{4+}$ ions with Ni increases the probability of double occupancy and therefore the on-site Coulomb interaction energy leading to a shift in the thermodynamically driven MIT to higher temperatures.

*Corresponding author
A. Bharathi, MSD, IGCAR,
Kalpakkam 603 102. India.
bharathi@igcar.ernet.in

## INTRODUCTION

It is now well established that the layered cobaltates, $Na_xCoO_2$, like the cuprates, are systems in which strong electron correlation plays a significant role in determining the ground state properties [1]. The role of Na is to donate electrons to the anti-ferromagnetic Mott insulator $CoO_2$. These electrons convert some of the $Co^{4+}$, sitting on a triangular lattice, to non- magnetic $Co^{3+}$. Channels of charge transport are thus created which induce metallicity in $Na_xCoO_2$. The initial interest in these compounds was the discovery of large thermopower in a metallic material [2]. The associated anomalous transport properties observed in this material, hinted at the role of strong electron correlations, in determining the physical properties. Early band structure calculations in $NaCo_2O_4$ [3] pointed out that the electronic states at the Fermi level would be the narrow 3d band of Co ions and that the large thermopower could arise due to the small Fermi surface. The interest in the material was further accentuated following the discovery of superconductivity at ~4.5 K in the water intercalated $Na_xCoO_2$ [4] for x=0.3. Several experiments followed, from which the exotic properties of the $Na_xCoO_2$ were brought to light. It is now clear that water intercalated $Na_{0.3}CoO_2$ is superconducting with a $T_C$ of 4.5 K, and that the un-intercalated compound, behaves like a correlated metal [5,6]. The compound with larger Na content (- 0.75), shows interesting magnetic anomalies at low temperature, has a sublinear temperature dependence of resistivity, and possesses the largest room temperature thermopower [ 2,7 ]. Further experiments [8] reveal that the thermopower is strongly suppressed in the presence of a longitudinal magnetic field bringing out the important role of spin degrees of freedom in determining this physical quantity. The large thermopower arises on account of entropy gained in the transport of an electron from the low spin $Co^{3+}$ state to a doubly (spin) degenerate, low spin $Co^{4+}$ state [1,9].

Thus the relative ratio of the $Co^{3+}/Co^{4+}$ is important in determining the ground state properties of $Na_xCoO_2$ and any substitution that brings about a change in this ratio could affect the magnitude of the thermopower and the resistivity behavior. This ratio can be altered by changing the Na content, and in fact it has been shown that the room temperature thermopower decreases with decrease in Na content [10], which has been reconciled within the spin entropy picture. The resistivity, on the other hand, increases with Na content [5,11].

Alternatively, the $Co^{3+}/Co^{4+}$ ratio could be altered by substituting the Co sublattice with other 3d transition metal ions. Very few reports of such substitution studies are available. Zhang et al [12] report a Metal Insulator Transition (MIT) at low temperatures upon replacing Co partially with Mn. This has been attributed to arise on account of disorder. Terasaki et al [7] find metallicity is retained for Cu substitutions up to 10 at% in place of Co. It should however be noted that the undoped sample, $NaCo_2O_4$, is reported to be metallic in this study, in contrast to the charge ordered insulating state reported recently [5]. Resistivity behavior reported by Yokoi et al [13] indicates a metal insulator transition for 2 at% Ir doping in $Na_{0.75}CoO_2$ at ~100 K. The samples remain metallic in the case of Ga doping up to 5 at% [13]. The results reported on Ir substituted samples [13] differ from batch to batch. Further,Yokoi et al [13] have concentrated on the effect of doping on the superconductivity of this system and have made no comments on the resistivity behaviour of the doped $Na_{0.75}CoO_2$ samples. We have studied the effect of Ni substitution on the $Na_xCoO_2$ samples and have examined them from the transport property measurements viz., resistivity in the 4 K-300 K temperature range. Infrared (IR) absorption measurements have been carried out to look for any local changes in the structure. The Co-O vibration mode has been followed as a function of Ni content. The temperature variation of these modes for the 2 at% Ni substituted sample has been studied for certain select temperatures across the metal insulator transition. In this paper, after briefly outlining the experimental details, we present the results and discuss these in the light of thermodynamic models proposed to understand metal insulator transitions [14,15].

## EXPERIMENTAL DETAILS

The samples used in this study were prepared by conventional solid state reaction technique using $Na_2CO_3$, $Co_3O_4$ and NiO as starting materials. The raw materials were first calcined at $600^0$C for 12 hours and subsequently pelletised and sintered at $850^o$C for 18 hours. To improve the grain connectivity which determine the resistivity and the thermoelectric properties, the samples were reground, pelletised and heat treated at $850^o$C in flowing oxygen for a further 20 hrs. X-Ray diffraction (XRD) measurements were carried out in a STOE diffractometer operating with $CuK_{\alpha}$ radiation in the Bragg-Brentano geometry. The Ni content in the samples was determined using a Philips ESEMXL30 Scanning Electron Microscope (SEM) equipped with an EDAX attachment. The compositions determined from these EDAX measurements were 15 at% for the nominal Ni, y=0.20, 5.24 at% for nominal y=0.05, and 10.63 at% for nominal y=0.10. From this analysis it is clear that for Ni fractions less than y=0.20, the actual Ni content is very close to that of the nominal composition. The sample with nominal y=0.20 is henceforth treated as the y=0.15 sample. The Na stoichiometry was determined by flame photometry. The sodium content determined in the pristine sample was $x=0.72\pm0.05$, whereas in samples containing Ni fraction, y=0.10 and y=0.15 it was seen to be $0.80\pm0.05$ implying that within measurement error the Na content remains same in all the samples irrespective of the Ni concentration. Four probe resistance measurements were carried out in a dipstick cryostat in the 4.2 K to 300 K temperature range. The PC data acquisition was automated using Labview 6.0 software. The room temperature resistivity was measured using van der Pauw geometry and the temperature dependent data were scaled using this value. Room temperature IR measurements were carried out in a BOMEM make DA8 FTIR spectrometer operating with a resolution of 4 $cm^{-1}$ in the far infra red range (30 $cm^{-1}$ to 900 $cm^{-1}$), using an extended 6 micron beam splitter and a DTGS detector, on samples pelletised with CsI. Low temperature measurements were carried out on samples, mounted inside a JANIS make continuous flow cryostat in the temperature range 4 K-300 K.

## RESULTS

The room temperature XRD patterns for $Na_xCo_{1-y}Ni_yO_2$ for various Ni fractions, y=0.0 to y=0.15 are shown in Figure 1. It can be seen from the figure that the samples are single phase with lines corresponding to the $\gamma$-$Na_xCoO_2$ phase, which crystallizes in the space group $P6_3/mmc$. A small impurity peak of NiO was seen in the diffractogram of the sample containing a Ni fraction of y=0.15 (shown in fig. 1). The fraction of NiO in this sample is estimated from PCW [16] program to be ~5%, in concurrence with the results from EDAX. The variations of the a-lattice and c-lattice parameters as a function of Ni fraction obtained by fitting the XRD line positions to the $P6_3/mmc$ structure, using the STOE program, are shown in Fig.2a. It can be seen from the figure that the lattice parameters show a non-monotonic variation with Ni fraction substituted. In particular the c-lattice parameter decreases from $10.901\pm0.004$ $A^0$ in the pristine sample to 10.821 $\pm0.004$ $A^0$ in the 3 at% Ni sample and further increases to $10.910\pm0.004$ $A^0$ for the 15 at% Ni sample, a value close to that in the sample with y=0. The behavior of the a-lattice parameter on the other hand is opposite to the behavior of the c-lattice parameter: it increases from $2.8301\pm0.0008$ $A^0$ in the pristine sample to $2.8374\pm0.0008$ $A^0$ in the 3 at% Ni substituted sample, beyond which it again shows a decrease attaining a value of $2.8277\pm0.0008$ $A^0$ for y=0.15. The variation of the cell volume with Ni content is shown in Fig.2b. For all Ni substituted samples up to concentration of 5 at%, the cell volume is lowered: it recovers to the value observed in pristine sample for y=0.10 and y=0.15.

The temperature dependence of the resistivity ρ(T) is plotted for the various Ni fractions in Fig.3. The van der Pauw resistivity measured at 300 K in the pristine sample is 8.0 mΩ-cm, which reduces with Ni substitution of 1 at% to 3.7 mΩ-cm. With further increase in the Ni fraction the resistivity increases reaching a value of 49.5 mΩ-cm, for the y=0.15 sample. The room temperature resistivity for Ni substitutions up to y=0.03 is smaller than that observed in the pristine sample. A systematic change in the ρ(T) with increase in Ni concentration is seen in Fig.3. The temperature dependence of the ρ(T) in the pristine sample is very similar to that reported in literature [17]. For Ni substitutions up to y=0.10 the resistivity curves show a positive temperature coefficient as the temperature is decreased below 300 K; there is a change in the sign of dρ/dT (becoming negative) at some lower temperature. The temperature at which this crossover occurs, is obtained from the derivative analysis of the temperature dependent resistivity and is termed $T_{mit}$. The variation of this temperature with increase in Ni content is shown in inset of Fig.3, from which it is clear that $T_{mit}$ increases with increase in y. For the sample with y=0.15 the resistivity shows a negative temperature co-efficient for the entire temperature range of investigation, implying that $T_{mit}$ would occur, if at all, at a temperature higher than 300 K.

In the temperature region where $d\rho/dT$ is negative and the sample shows a semiconducting like behavior, plots of log($\rho$) versus various powers of 1/T have been attempted and it was found that best fits were obtained for $1/T^{1/2}$. The plots of logarithm of resistivity versus $1/T^{1/2}$ are shown in Fig.4. The data on the sample with nominal y=0.15 is a straight line in the entire range of temperature investigated. For smaller concentrations the $1/T^{1/2}$ behavior is seen to be valid in the insulating regime (cf. Fig.4). From Fig.4 it can be concluded that the resistivity behavior in the insulating regime conforms to the temperature dependence dictated by variable range hopping transport of the Effros-Shklovskii type, viz., $\rho(T)= C\exp-(T_0/T)^{\alpha}$, where C=constant and $T_0$ the variable range hopping parameter, with the exponent $\alpha$=0.5 [18].

In the pristine sample the resistivity behaviour fits to $T^{\beta}$ where $\beta$=0.47-0.5 in the low and high temperature regimes respectively. Similar sub-linear temperature dependence of resistivity has been described by Bruhwiler et.al [17] for different batches of single crystalline and polycrystalline samples. In contrast, in the metallic regime of all the Ni substituted samples, the temperature dependence of resistivity is well described by a $\rho(T) = \rho_0+AT^2$ as shown in Fig.5 for Ni fractions up to y=0.05. The solid lines (cf. Fig.5) correspond to linear fits from which the parameters $\rho_0$ and A are extracted and are plotted in Fig.6 as a function of Ni content. It can be seen from the figure that $\rho_0$ shows an increase with Ni substitution. The coefficient of the $T^2$ term, A, increases upto 4 at% Ni substitution, beyond which it decreases for the 5 at% sample. The observation of $T^2$ dependence of resistivity implies that electron-electron scattering is dominant in determining the transport in the Ni substituted samples. In several systems undergoing metal insulator transitions, it has been shown that A increases as a consequence of an increase in the effective mass of electrons, primarily arising from increased interactions as electrons tend to localize when the system approaches the insulating state [19]. From Fig.6 it can also be surmised that the magnitude of the resistivity contributions from both the constant term ($\rho_0$) and the $T^2$ term are of similar magnitude at high temperatures, whereas at low temperatures the residual resistivity term, $\rho_0$ tends to dominate. This is in contrast to the resistivity behaviour seen in the pristine system, where there is no perceptible residual resistivity term [17] and spin fluctuations dominate temperature dependence of resistivity [20]. This behaviour hints at a complete change in the nature of the ground state of the system with Ni substitution. A similar change in the temperature dependence of resistivity behaviour has also been seen by Foo et al [5], in which a decrease in the Na stoichiometry results in the dominance of the resistivity by the $T^2$ term in the metallic phase of the samples. We now focus on the observed variation of $\rho_0$ and A as a function Ni fraction substituted. The variation in $\rho_0$ with Ni substitution shows an initial increase at 785 μΩ-cm/at%. This variation is large when considered in the context of transition metal oxides [19]. Further it can be seen that between 3 at% and 4 at% there is a jump in $\rho_0$ by 4600 μΩ-cm. The residual resistivity variation is given by [19]

$$\rho_0 = \left\{ \frac{e^2 S_F L_i}{12\pi^3} \right\}^{-1} \qquad \text{------------------------(1)}$$

where $S_F$ is the mean Fermi surface area and $L_i$ is the mean impurity distance, which is inversely proportional to square root of the concentration of Ni. This implies that $\rho_0$ should increase as a square root of the Ni concentration ( see dashed line Fig. 6) if $L_i$ was the only factor that influences the increase of $\rho_0$. In contrast, it can be seen from Fig. 6a that there is a jump in $\rho_0$ between 3at% Ni and 4 at% Ni sample, deviating distinctly from the behaviour expected from the increase in Ni concentration. From eqn. (1) the jump in $\rho_0$ can be attributed to a sudden decrease in the Fermi surface area at Ni concentration of 4 at%, indicating a shrinkage of the Fermi surface or a reduction in carrier density at this concentration. The A parameter shows a small increase with increase in Ni content upto 3 at%, but at 4 at% it shows a distinct increase, before decreasing in the 5 at% sample. The behaviour of A and $\rho_0$ as a function of Ni fraction is similar to the behaviour seen in the $NiS_{2-x}Se_x$ system when the system goes over from the paramagnetic metallic to a paramagnetic insulating phase [19]. The increase in the A parameter with Ni substitution, indicates an increase in the mass re-normalisation effect, since $A \sim \gamma^2$; where $\gamma=\gamma_0(m^*/m_0)$ is the coefficient of the linear term in specific heat and $m^*$, the effective mass in the presence of electron-electron interaction and $\gamma_0$ and $m_0$ are the parameters in the absence of electron-electron interactions [15,19]. It should be emphasized that at 4 at% Ni concentration, A shows a maximum (see Fig.6b) and an anomaly is also seen in $\rho_0$ at this composition (cf. Fig. 6a), both of which point to strong mass renormalization at this concentration.

To obtain further insight into the occurrence of the MIT, far IR measurements were carried out for the entire series of $Na_xCo_{1-y}Ni_yO_2$ samples. The spectrum of the pristine sample reveals a low frequency mode at 260 $cm^{-1}$ corresponding to the in-plane vibration of Na and a broad high frequency feature comprising both of the in plane and out of plane Co-O vibrations. This spectrum shown in the topmost panel Fig. 7a, is similar to that reported earlier [21,22]. The room temperature IR spectra in all Ni substituted samples are compared in Fig.7a. It is evident from the figure that no new modes appear with Ni substitution. Although the IR mode corresponding to Na does not shift with Ni substitution, the modes due to Co-O bonds show significant changes in line shape. While the low frequency Co-O mode is relatively unaffected and has been fitted to a Lorentzian the high frequency in plane mode is seen to develop an asymmetry with increased Ni content. Such an asymmetry towards high frequencies points to an interaction of this phonon mode with a continuum of states. Hence this mode is fitted to a Fano line shape [23,24] given by

$$I = I_0 \frac{[1 + \{\omega - \omega_0\} / \Gamma q]^2}{1 + [\{\omega - \omega_0\} / \Gamma]^2} \qquad \text{----------------------(2)}$$

Where $\omega$ is the renormalized phonon frequency, $\omega_0$ is the bare phonon frequency, 1/q is the Fano asymmetry parameter, a measure of the strength of the coupling between a discrete mode and the continuum and $\Gamma$, the width of the resonance interference between the continuum and the discrete channels. Only the in plane Co-O mode couples with the electronic system, as the electron transport occurs between $Co^{3+}/Co^{4+}$ predominantly along the ab plane in this compound [1]. The fits to sum of Lorentzians the latter modified by a Fano factor is shown for the room temperature data for the Co-O vibrational modes in Fig. 7b. The fits are fairly satisfactory. The extracted Fano parameter 1/q as a function of Ni concentration y, initially increases with Ni concentration, reaching a maximum for y=0.02, after which the mode tends to become more symmetric, implying that at 300 K the relevant excitations in the continuum couple more strongly for the low concentration Ni substituted samples. The variation in the Fano asymmetry parameter of the Co-O mode at representative temperatures is shown in Fig. 8 for the 2 at% Ni substituted sample, along with the corresponding variation in the mode frequency and line width. It may be noted that 1/q which is a measure of the interaction strength between the Co-O mode and the continuum grows in magnitude as the temperature is lowered reaching a maximum value close to the MIT. Below 50 K a distinct increase in the mode frequency to 600 $cm^{-1}$ occurs, the temperature at which the sample shows an insulating behaviour (cf. Fig.3). The line width $\Gamma$ variation shown in Fig.8c, indicates a small gradual increase with decrease in temperature.

## DISCUSSION

From the results presented above two important facts emerge; At the lowest temperature of measurement viz., 4.2 K the system shows a drastic change from a metallic behaviour in the pristine sample to an insulating state for the smallest Ni concentration studied. Second, the temperature dependence of resistivity shows a qualitative change from a $T^{0.5}$ behaviour in the pristine sample to the $\rho_0 + AT^2$ behaviour in the metallic region in Ni substituted samples. Such an unusual changeover can arise only if Ni induces a significant change in the electronic structure.

It is now well known in literature that the $Na_xCoO_2$ system is unstable to charge ordering for various compositions of x [6], which has been experimentally verified for the case of x=0.5 [5]. In $Na_xCoO_2$ there is only fractional filling of charge carriers since the fraction of $Co^{4+}$ ions in the $d^5$ configuration with one hole per site is only (1-x). Hence the insulating state observed in this system cannot be of the conventional Mott Hubbard type which requires integral filling of charge carriers per site. If the fraction x is rational (like 0.75) then the available $Co^{4+}$ ions can order in a manner commensurate with the parent lattice. Such a spatial ordering of the charges can lower the Coulomb interaction energy in the system [25]. Any excitation from this charge ordered state will have a finite gap. We show below that an insulating state is stabilized at the lowest temperatures in Ni substituted samples. This insulating phase could be a charge ordered state. Charge transport in this charge ordered insulating state can occur, whenever thermally induced excitations relaxes this order, by hopping of holes from Co $3d^5$ to Co (Ni) $3d^6$. Since this transport takes place between highly correlated atomic 3d states and there is a distribution of distances over which this hopping occurs one obtains a correlated variable range hopping transport observed experimentally (see Fig. 4). With increase in temperature the charge ordered insulating state transforms to a correlated metallic state with the conventionally observed, $\rho_0 + AT^2$ temperature dependence of resistivity. The variation in $\rho_0$ and A with Ni concentration (see Fig.6) clearly suggests the tendency of the charge carriers to localize indicative of the preference of an insulating ground state in Ni rich samples. It may be pointed out that the unusual temperature dependence in the pristine $Na_{0.75}CoO_2$ sample has not been clearly understood and has been variedly thought to arise due to the system being a Curie Weiss metal [5] or a quantum charge liquid [26]. Without a clear understanding of the metallic behavior of $Na_{0.75}CoO_2$ and the scattering mechanisms that go to determine the evolution of resistivity with temperature it is not possible to comment on the sudden reduction of $\rho_{300K}$ with 1 at% Ni addition, excepting to attribute it to a change in the nature of the ground state. The further increase in resistivity with increase in Ni fraction could be due to increase in residual resistivity as clearly seen from Fig. 6a.

To understand the occurrence of MIT as a function of temperature and the Ni concentration dependence of the $T_{mit}$, we follow the thermodynamic arguments given by Spalek [14,15], to explain the MIT in $(V,Cr)_2O_3$. In the system we have investigated, the metal insulator transition is associated with the delocalisation-localisation of the 3d electrons in the $Co^{4+}$ sub-lattice. The transition can be viewed as a thermodynamic transition from the metallic phase to an insulating phase wherein the charge and the spin are confined at the site of $Co^{4+}$ ions. The free energy of the system [14,15] in the metallic state per particle, $F_M/N$ is given by

$$\frac{F_M}{N} = -\varphi bW + U\eta - \frac{\gamma_0 T^2}{\varphi} \qquad \text{-------------- (3)}$$

Where $-\varphi bW$ is the reduction in the free energy due to de-localisation of holes, W being the band width and $\varphi$ the band narrowing factor describing the degree of difficulty for the particles to hop from one atom to another in forming a band state. b takes into account the fractional filling of charge carriers per atomic site; which decreases with increase in y. The term $U\eta$, describes the increase in Free energy due to strong electron correlation effects; where U is the on-site repulsion energy and $\eta$ is the probability of double occupancy at an atomic site (see below). The third term in Eq. (3) arises from the reduction in free energy brought about by the smearing of the Fermi distribution at finite temperature taken in the low T limit [14] and $\gamma_0$ as defined earlier is the coefficient of the linear term in the electronic specific heat in a non interacting electron gas.

Before proceeding further it is better to clarify the meaning of U and $\eta$ in the context of the system $Na_xCoO_2$ under discussion (see level diagrams in Fig.9). We assume that it costs an energy U to convert a Co $3d^5$ to $3d^4$ (double occupancy by holes), or $3d^5$ to $3d^6$ (double occupancy by electrons) and Ni $3d^6$ to $3d^5$ (double occupancy by holes, since Ni $3d^5$ already has a hole residing at its site). In the pristine sample (y=0.0) electrical conduction occurs by the hopping of a hole from $Co^{4+}$ ($3d^5$) to $Co^{3+}$ ( $3d^6$). This transfer conserves the number of Co ions in the $3d^5$ and $3d^6$ configurations and does not result in any additional double occupancy and therefore does not cost any U. Transfer of a hole from a $Co^{4+}$ to another $Co^{4+}$ ion would leave the former in a $3d^4$ and the latter in a $3d^6$ configuration. The occurrence of Co in $3d^4$ state is very rare and if it does occur then a high spin state would be the preferred option. No evidence for this has been found. Further such a transfer would cost an energy 2U, since, the final state would have a doubly occupied hole ($3d^4$) at one site and an extra doubly occupied electron ($3d^6$) at another site. Both these sites were singly occupied before the transfer. Hence we rule out the possibility of a hole transfer into a Co ion in $3d^5$ state from another Co ion in the same state. Under such circumstances, in the pristine sample, the U term in Eq. (3) above need not be considered. This statement should not be taken to mean that we are ignoring correlations altogether in $Na_xCoO_2$. We are only asserting that charge (hole) transport can occur without any punitive U cost in energy via the $Co^{4+} \rightarrow Co^{3+}$ channel. Hole or electron transport via the $Co^{4+} \rightarrow Co^{4+}$ channel is forbidden because of the 2U cost in energy. Upon Ni substitution ( Ni enters the lattice in $3d^6$ configuration and has a hole residing at its site) the transfer of a hole from a Co $3d^5$ to Ni $3d^6$ would result in a double hole occupancy at the Ni site and would therefore cost an energy U. No additional electron double occupancy occurs because of this transfer. The probability, $\eta$, of this double hole occupancy would increase with Ni concentration y, which would lead to an increase in the band narrowing factor $\varphi$, and would lead to a narrower band. Further, Ni substitution would deplete the system of $Co^{4+}$ ions resulting in reduction in the value of b and these would result in a decrease in the first term in Eq.(3). The net result would be an increase in the value of the temperature independent factor $U\eta$ - $\varphi bW$ in Eq. (3) above, which goes from a negative value for U=0 ( in pristine sample) to increasingly positive values as y is increased. These are schematically shown in Fig.9, which is a plot of $F_M/N$, $F_I/N$ versus the temperature, T. It is evident from the figure that the intercept on the Free energy axis is a measure of the $U\eta$ - $\varphi bW$.

In the insulating state the free energy per particle is given by

$$\frac{F_I}{N} = -k_B T \ln 2 \qquad \text{----------------(4)}$$

which arises due to the spin entropy associated with the spin half charge carriers at the $Co^{4+}$ sites, that lowers the free energy of the system with increase in temperature. When the two energies (Eq.3 and Eq.4) match, the system becomes unstable towards a phase transition. It can be seen from Fig. 9 that at T=0, the free energy in the insulating phase is smaller than that in the metallic phase, when the intercept is positive. Whereas when the intercept is negative the free energy of the metallic phase is lower than that of the insulating phase at T=0 and in the particular free energy diagram sketched (cf. Fig.9a), the metallic phase is stabilized in the entire temperature range. An increase in $\eta$, as happens with Ni substitution, can result in the intercept becoming positive, in which case at low temperatures the insulating state can be stabilized as shown by Fig.9b, and 9c. As the temperature rises the free energy reduction resulting from the $T^2$ term in Eq. (3) compensates for the increase in $U\eta$ term and at the temperature at which $F_M$ crosses $F_I$ a transition to the metallic state occurs (cf. Fig.9b and 9c). The drastic change in the sign and magnitude of the intercept along the free energy axis which comes about because of the switching on of U from zero to a finite value upon substituting $Co^{4+}$ in $3d^5$ state by $Ni^{4+}$ in $3d^6$ state, could result in the metal-insulator transition with the smallest Ni substitution. As can be clearly seen from Fig. 9 the metal insulator transition occurs at a higher temperatures in samples containing a higher fraction of Ni. A striking experimental evidence for the build up of the coulomb correlation with Ni substitution is the observation of an increase in the A co-efficient of the $T^2$ term in the temperature dependent resistivity, with increase in Ni content (see Fig.6b). It is worth pointing out that in $Na_xCoO_2$, the addition of Na and substitution of Ni is not the same; as Na converts a Co ion from a $3d^5$ to $3d^6$ with no holes, whereas Ni addition converts $Co^{4+}$ to $Ni^{4+}$ viz., $3d^5$ to $3d^6$ with a hole at its site.

As already mentioned in the results section, the asymmetry in the phonon lineshape arising due to Fano resonance is caused by the interaction of the vibrational degrees of freedom with some elementary excitations in the electronic continuum. These excitations can be in the form of polarons, electron-hole pairs (excitons) etc[27,28]. Since the Ni substituted sodium cobaltates become insulating at low temperatures, the plausible excitations could be electron-hole excitations. As the temperature is lowered, the IR results on the 2 at% Ni substituted sample (cf. Fig.8a) clearly indicates a build up of this interaction strength as one approaches the MIT. This could arise due to a higher probability of the creation of the excitons just above the MIT. The decrease in the value of 1/q below $T_{mit}$ could arise due to the freezing of the electron-hole excitations in the insulating phase. This decrease in the coupling strength could also be responsible for the observed increase in the mode frequency (see Fig. 8b) below $T_{mit}$. The decrease in the room temperature value of 1/q for Ni additions greater than 2 at% points to a weakening of the interaction strength, which obtains as a consequence of the increase in the insulating gap with Ni content and the Fano resonance drifting away from the phonon mode.

As discussed in the foregoing paragraph, in the metallic phase of the Ni substituted samples IR results show an increase in the phonon-continuum interaction strength. These, plus the dominance of the electron-electron scattering strength in determining the temperature dependence of resistivity (cf. Fig.6) imply that the elementary excitations responsible for the Fano resonance do not take part in electrical conduction.

It is interesting to note that the room temperature a- and c-lattice parameters extracted from XRD measurements (cf. Fig.2a) and the corresponding cell volume (cf. Fig. 2b), exhibit a non-monotonic behaviour with Ni addition. In particular as seen from Fig.2b, the lattice volume, measured at 300 K, is lower in all Ni substituted samples that are metallic at this temperature, having undergone a MIT at some lower temperature. In the y=0.15 sample (insulating at 300 K) and the y=0.10 sample (on the verge of becoming an insulator at 300 K), there is no perceptible lattice compression observed. This compression of the lattice in the metallic phase of the Ni doped samples could occur because the system can gain in kinetic energy by the delocalization of the charge carriers [29]. With further increase in Ni content no compression is observed as the insulating phase is the stable phase at the temperature of the XRD measurement. In other words, Ni substitution increases the free energy of the system on account of an increase in the Coulomb repulsion; to compensate for this the system undergoes a compression, thereby gaining in delocalization energy. The cell volume decreases and the system turns more metallic. For large Ni concentration the Coulomb repulsion term dominates and the sample turns insulating.

## SUMMARY

To summarize, electrical resistivity and IR spectroscopy measurements have been carried out in $Na_xCo_{1-y}Ni_yO_2$ samples as a function of temperature in the 4 K-300 K range for y up to 0.15. The smallest addition of Ni changes the ground state to an insulating one. As the temperature is raised the system becomes metallic giving rise to a thermodynamic metal insulator transition for each Ni concentration. The temperature at which the MIT occurs for each of the Ni concentration is obtained from a competition between the free energy in the metallic regime and that in the localized regime. It turns out that the increase in the correlation term due to the substitution of Ni for $Co^{4+}$ ions is largely responsible for the higher free energy in the metallic regime at T=0, and shifting of the $T_{mit}$ to higher temperatures. The signature of the occurrence of the MIT shows up in the variation of the Fano asymmetry parameter in the Co-O in-plane vibrational mode, both as a function of temperature and Ni content.

## ACKNOWLEDGEMENT

We thank Dr. Manas Sardar of Materials Science Division, IGCAR for valuable discussions. We also thank Dr. Saroja Saibaba of Physical Metallurgy Section, IGCAR for the EDAX measurements and Mr. A. Thiruvengadasami of Material Chemistry Division, IGCAR for the flame photometry measurements.

## FIGURE CAPTIONS

Fig. 1 The XRD patterns in $Na_xCo_{1-y}Ni_yO_2$ samples. Each data is normalized to the maximum peak height and the data shifted vertically for clarity.

Fig. 2 (a)Variation of the a-axis and c-axis lattice parameters as a function of Ni content. Solid lines are guides to the eye. (b) Variation of the unit cell volume as a function of Ni content.

Fig.3 Variation of ρ(T) with temperature for all Ni substitutions, y, indicated on the curves. Inset shows the $T_{mit}$ identified as the temperature at which dρ/dT changes sign, as a function of Ni fraction y.

Fig. 4. Log(ρ) versus $1/T^{1/2}$ in $Na_xCo_{1-y}Ni_yO_2$ for various nominal y, note (a) is for high Ni concentrations and (b) is for low concentrations. The curves pertain to the insulating phase of the samples.

Fig. 5 The variation in resistivity with $T^2$ in the metallic region. Solid lines are straight line fits, from which residual resistivity and the coefficient of the $T^2$ term are calculated.

Fig. 6 The variation of the parameters (a) $\rho_0$ and (b) A obtained from the fits to the temperature dependent resistivity in the metallic region, as a function of Ni content y; solid lines are guides to the eye. The dashed line in (a) is the expected variation of $\rho_0$ if only the variation in concentration was important.

Fig.7 (a) The room temperature IR absorption spectra in $Na_xCo_{1-y}Ni_yO_2$ samples in the 200 $cm^{-1}$ and 700 $cm^{-1}$ wavenumber range; the data have been shifted along the vertical axis for clarity. (b) Co-O mode behaviour at 300 K from IR absorption data (open circles) shown along with fits (thick solid line) to a Lorentian+ Lorentian Lineshape modified by Fano asymmetry parameter. The individual components of the fit are shown as thin lines.

Fig.8 The variation of (a) the Fano asymmetry parameter (1/q), (b) the mode frequency and (c) line width, as a function of temperature for sample containing 2 at% Ni.

Fig.9 Schematic diagram of the free energy of the insulating phase $F_I$ and that of the metallic phase $F_M$ as a function of temperature. (a) $F_M$ with a negative intercept (b) $F_M$ with a small positive intercept corresponding to a small addition of Ni and (c) same as (b) but with a larger intercept corresponding to a larger Ni addition. The level diagram for $Co^{3+}/Ni^{4+}$ in $3d^6$ configuration and $Co^{4+}$ in $3d^5$ configuration both in low spin are also shown.

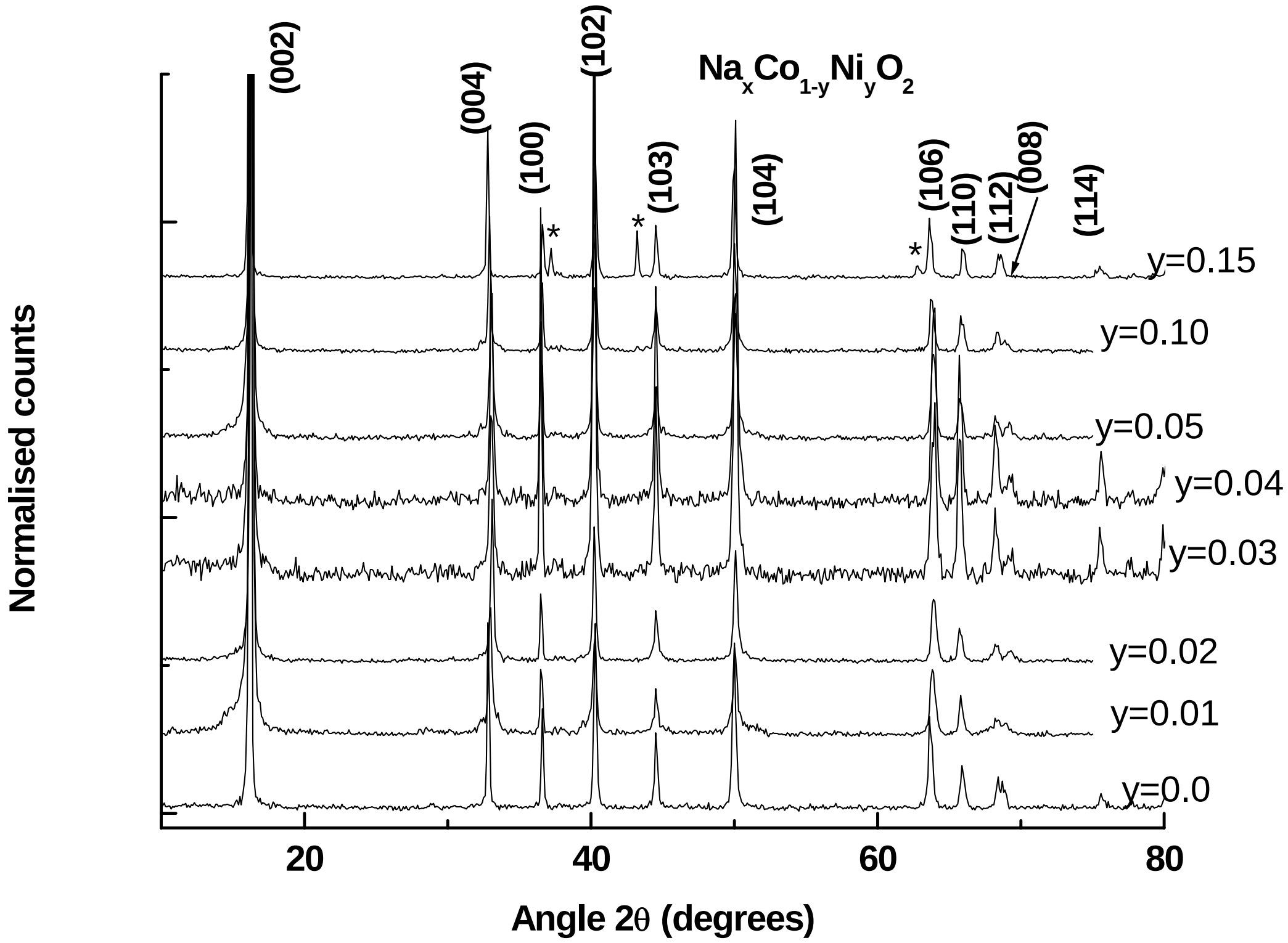

Fig. 1 Gayathri et. al

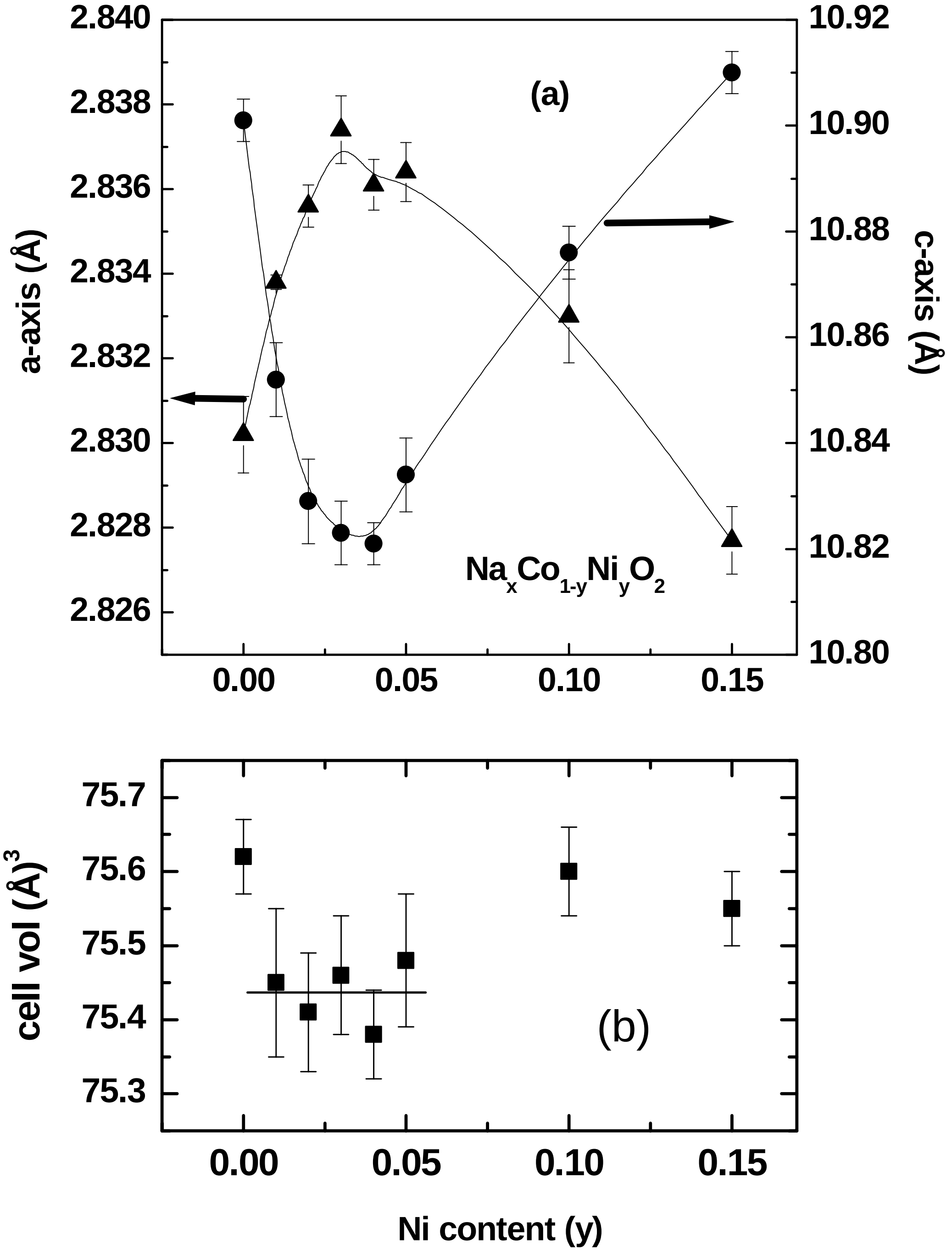

Fig.2 Gayathri et al

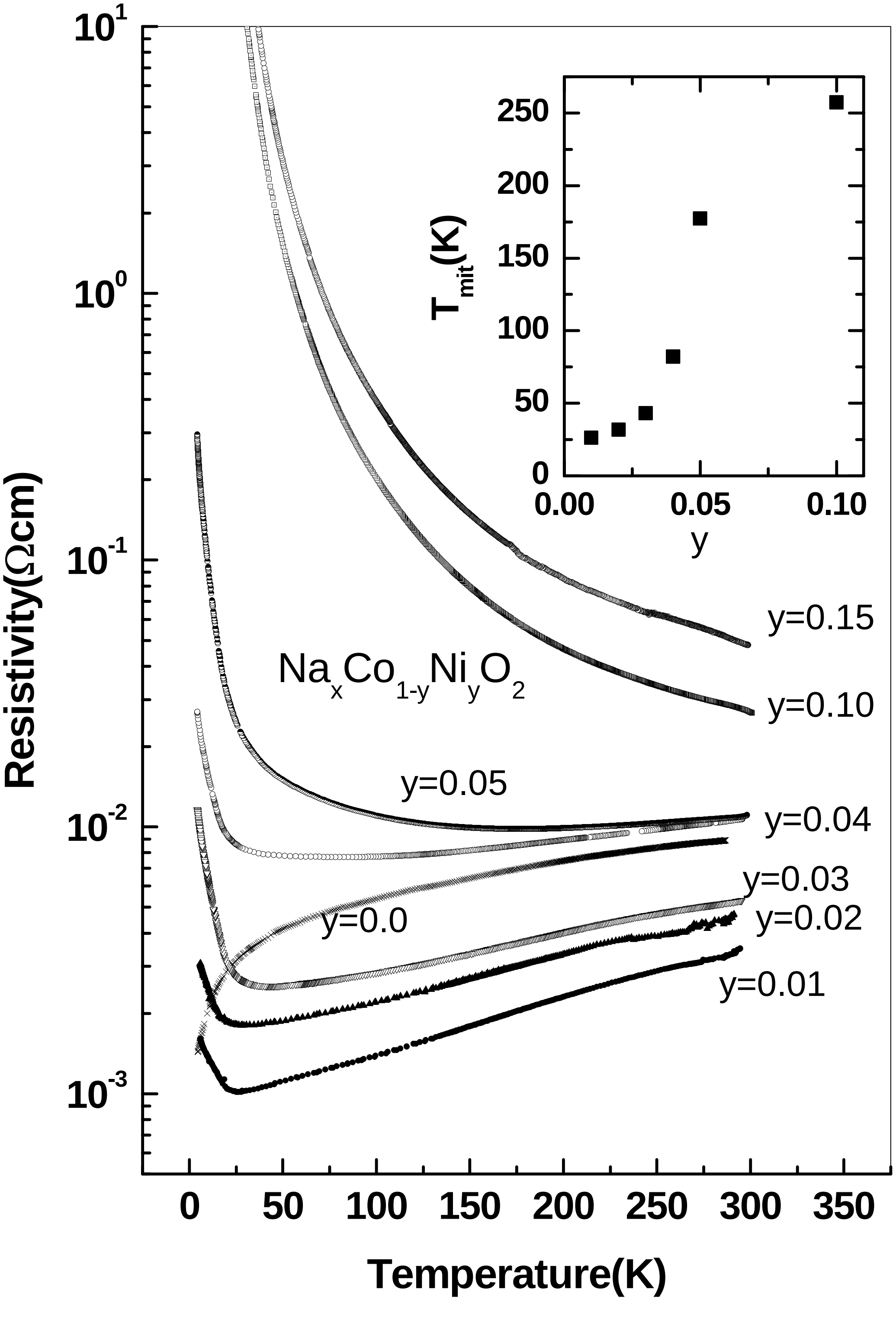

Fig.3 Gayathri et.al

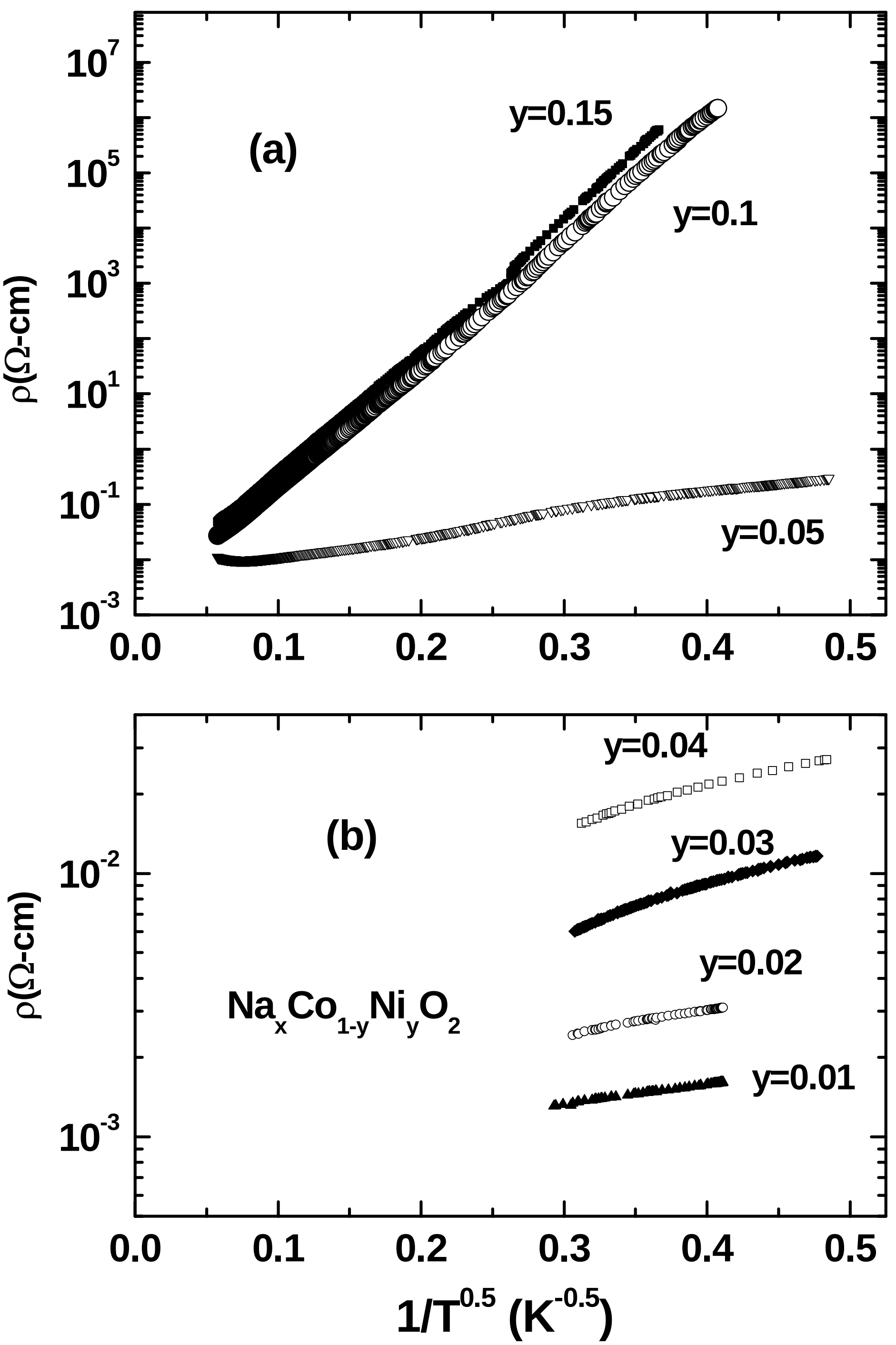

Fig.4. Gayathri et. al

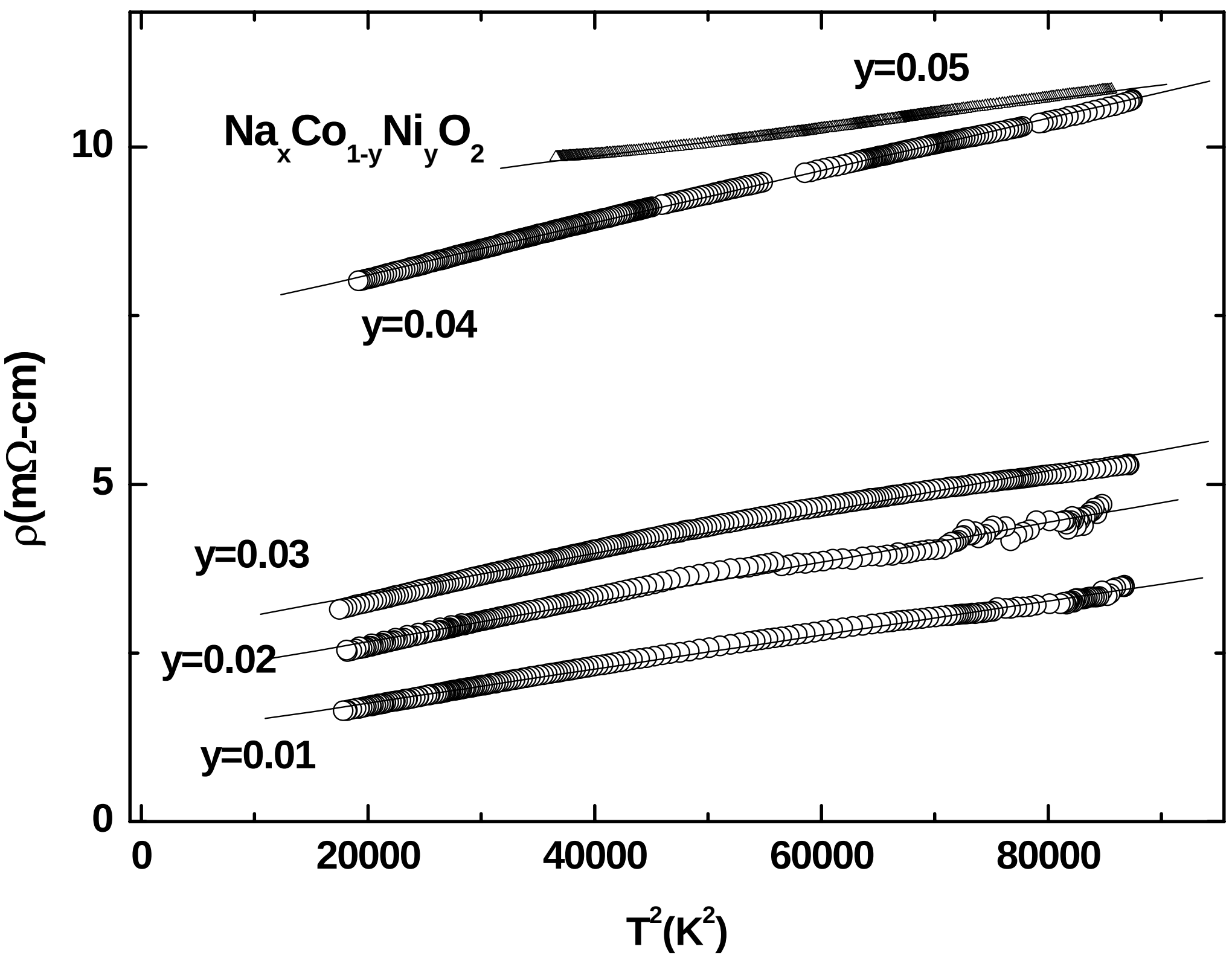

Fig. 5 Gayathri et al

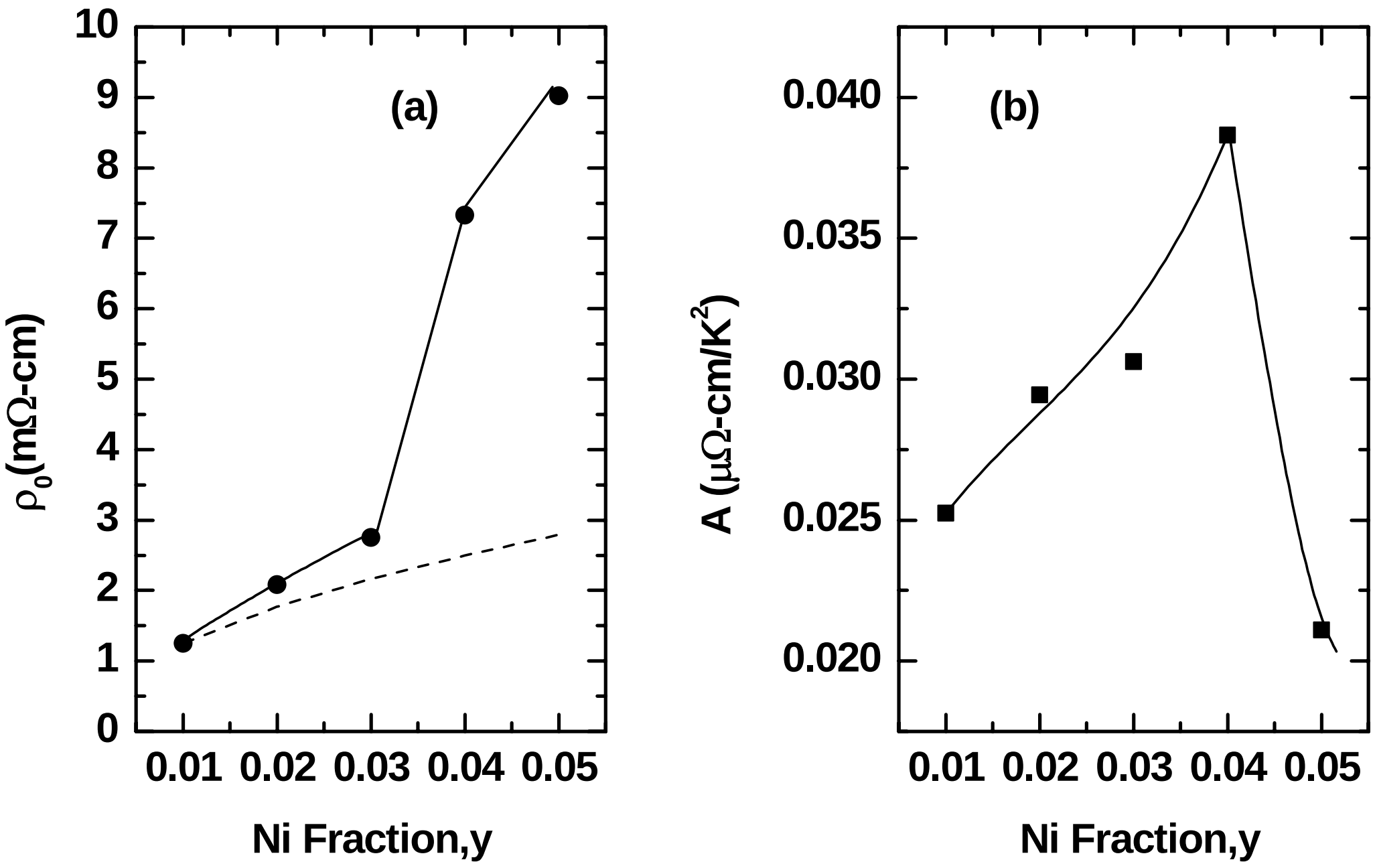

Fig.6 Gayathri et al

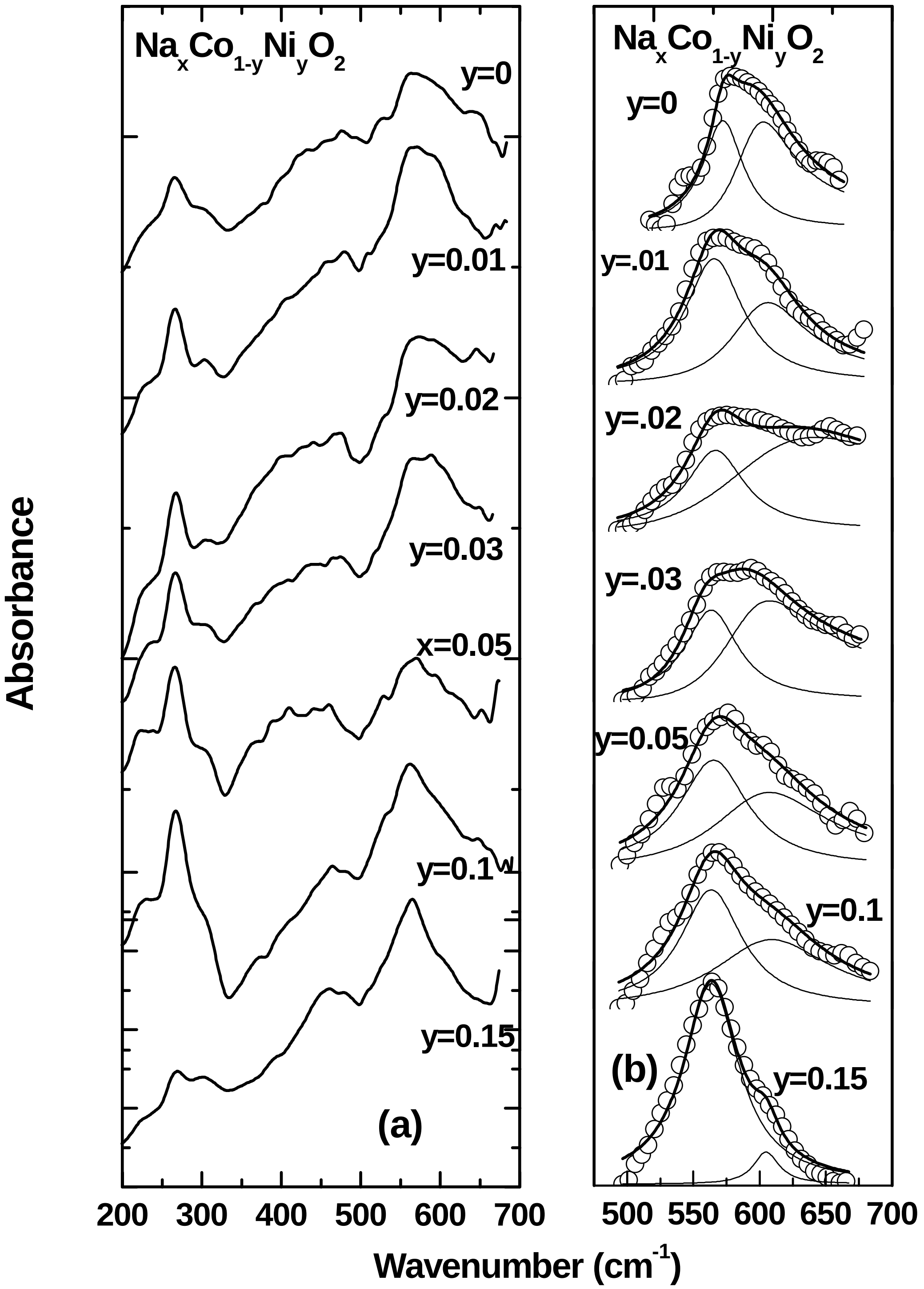

Fig.7 Gayathri et. al

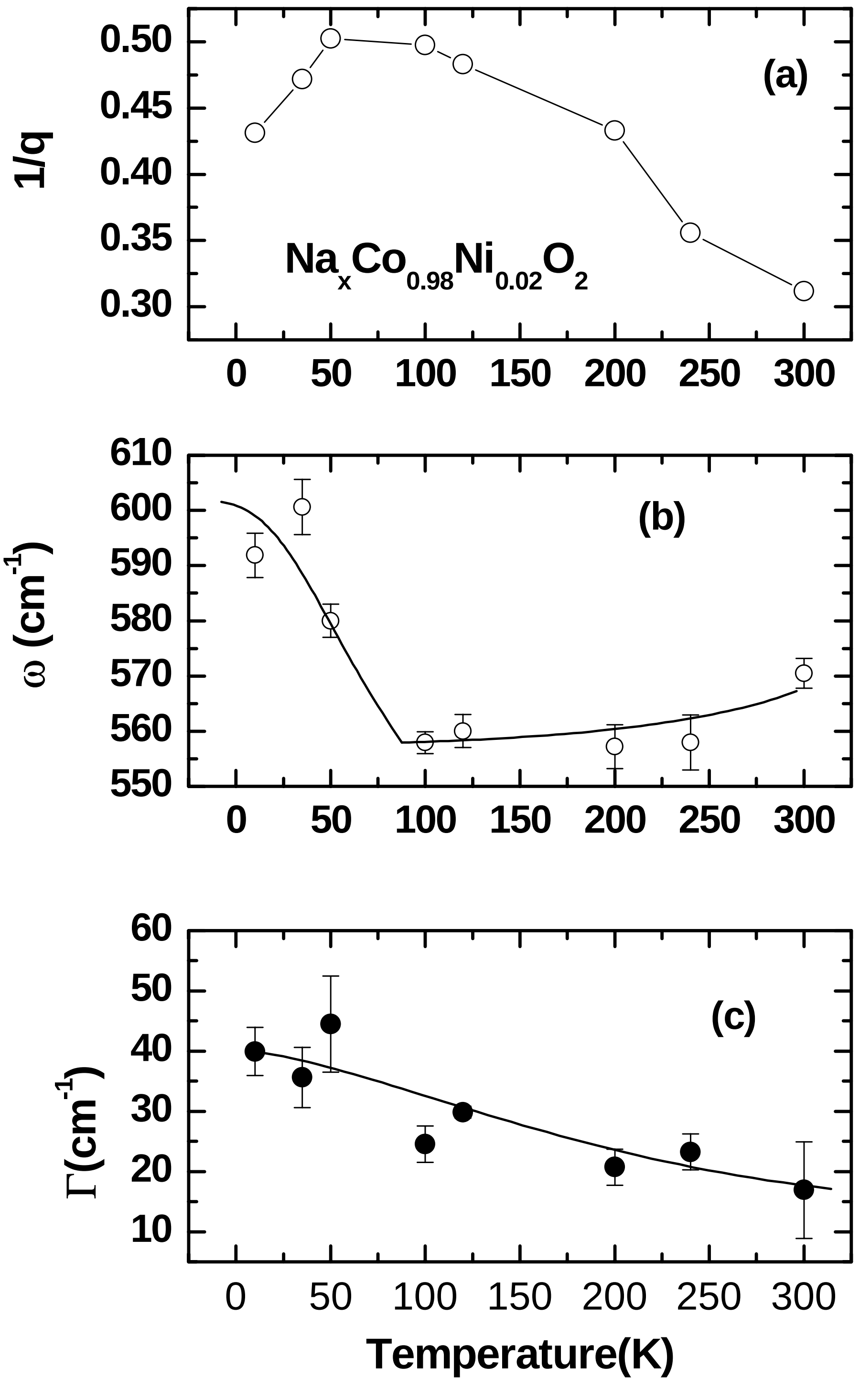

Fig. 8 Gayathri et. al

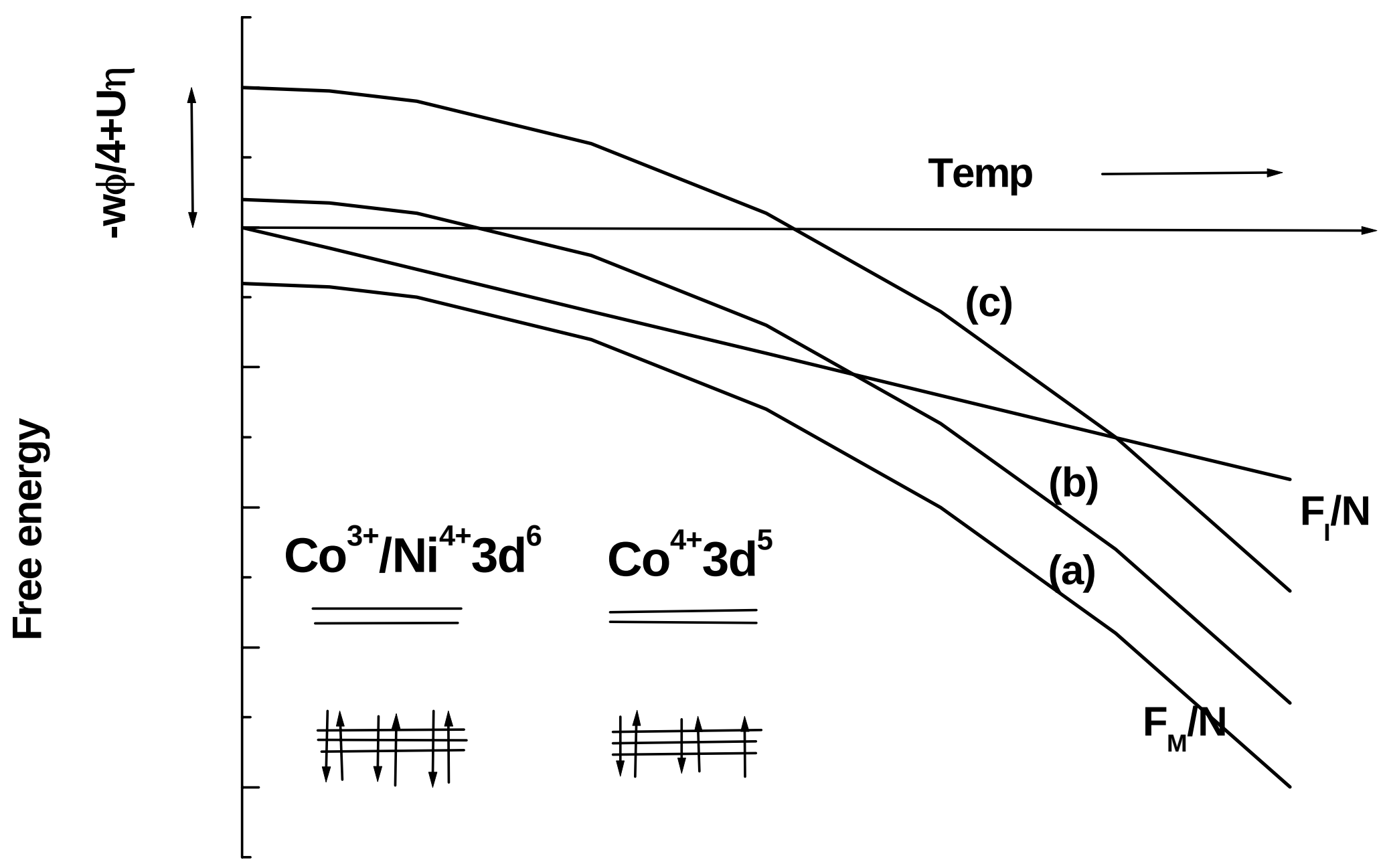

Fig.9 Gayathri et al